# Twist-induced near-field radiative thermal regulator assisted by cylindrical surface modes

Jian-You Wang [a, b], Yong Zhang [a, b, *], Xiao-Ping Luo [a, b], Mauro Antezza [c, d], Hong-Liang Yi [a, b]

[a] *School of Energy Science and Engineering, Harbin Institute of Technology, 92, West Dazhi Street, Harbin 150001, PR China*

[b] *Key Laboratory of Aerospace Thermophysics, Ministry of Industry and Information Technology, Harbin 150001, PR China*

[c] *Laboratoire Charles Coulomb (L2C) UMR 5221 CNRS-Université de Montpellier, Montpellier F-34095, France*

[d] *Institut Universitaire de France, 1 Rue Descartes, Paris Cedex 05 F-75231, France*

**ABSTRACT**

Near-field radiative heat transfer (RHT) can surpass Planck's blackbody limit by several orders of magnitude due to the tunneling effect of thermal photons. The ability to understand and regulate RHT is of great significance in contactless energy transfer. In this work, we construct a rotating system with a hexagonal boron nitride (*h*-BN) cylinder for actively regulating the RHT between two nanoparticles (NPs). The results show that when the two NPs are located directly above the cylinder, energy can be directionally transmitted along the cylindrical channel in the form of low-loss surface waves, which can significantly enhance RHT. In addition, we find that the RHT can be regulated by actively manipulating the excitation of cylindrical surface modes. When the rotation point is located in the middle of the line connecting the two NPs, the modulation contrast approaches five orders of magnitude, higher than that of cylinders made of other materials under the same conditions. When its diameter is slightly less than the distance between NPs, the *h*-BN cylinder shows excellent tunability in the heat exchange. The present work may offer a theoretical possibility for actively regulating and controlling near-field RHT between arbitrary objects based on cylindrical waveguides.

---

[*]Corresponding author: yong_zhang@hit.edu.cn (Y. Zhang)

# 1. Introduction

Photon-mediated radiative heat transfer (RHT) relies on electromagnetic modes to transport energy between two interacting objects at different temperatures [1-3]. When the spacing between the objects is comparable to or less than the thermal wavelength (around 7.6 μm at room temperature), the near-field RHT dominated by evanescent modes (photon tunneling) exceeds the blackbody limit by several orders of magnitude [4-9]. In recent years, rapid advances in thermal photonics, materials science, and quantum physics have promoted the development of near-field RHT, which has received widespread attention in many application fields, such as thermal management, optoelectronics, and energy-conversion devices [10-14]. In addition, the active control of energy transport plays a substantial role in maintaining the stable operation of equipment under different working conditions. Therefore, the ability to understand and regulate the near-field RHT is of great significance for achieving efficient energy transfer and utilization at the micro- and nanoscale.

Many scientists and scholars have theoretically and experimentally proposed some feasible strategies to enhance and actively regulate near-field RHT. These research efforts can be divided into several categories. The first category is to change specific material properties by applying external fields (electric, magnetic, and velocity fields), thus regulating the RHT between objects. Graphene is an ideal material for controlling near-field thermal radiation because the plasmonic resonance it exhibits in the mid-infrared region can be modified by introducing an electric field [15]. Also, by applying external magnetic fields to Weyl semimetals, magneto-optical, ferromagnetic, and phase-change materials, a variety of intriguing phenomena can be observed, such as the thermal magnetoresistance effect [16], the photon thermal hall effect [17,18], the magnetocaloric effect [19], and thermal rectification [20-24]. The RHT between objects can be effectively regulated. Based on the electrodynamics theory of moving media, some works have investigated the one-way hyperbolic propagation properties of metasurfaces after applying a velocity field [25,26]. These works provide a new method for controlling the near-field RHT between objects. The second category is based on many-body interactions to stimulate the coupling of evanescent surface modes, thus actively controlling the heat exchange [27]. The objects of study include particle chains and clusters [28-33], as well as multilayer planar structures [34-37]. Tremendous efforts have been made on structures such as bulks, single or multilayer films, spheres, and cylinders to enhance or regulate near-field RHT [38-44]. In addition, based on the concept of twistronics in the low-dimensional system, the rotational degree of freedom (also known as the rotation control) is another feasible strategy for actively controlling energy transport [45-47]. The RHT between objects can be regulated by mechanical rotation, which can precisely control the energy bands of the material. Recently, this method has attracted widespread interest in Casimir interactions and near-field RHT [48]. It is worth noting that in most research works, these categories may be presented simultaneously in a combined manner to obtain excellent enhancement and regulation performance.

According to the literature review, the studies on the regulation and active control of near-field RHT mainly focus on planar structures, while there are relatively few studies on the thermal transport properties of other structures. Recently, a perfect conducting cylinder analogous to a superconductor has been proposed, which can transfer electromagnetic energy over arbitrarily long distances with almost no loss [49]. It has been shown that such cylinders can significantly enhance the RHT between two objects over long distances [50-52]. Inspired by the excellent energy transport properties of cylindrical structures and the concept of twistronics, we propose a twist-induced near-field radiative thermal regulator consisting of two nanoparticles (NPs) and a cylinder. The heat exchange between the two NPs can be regulated by actively manipulating the excitation of cylindrical surface modes. We describe the mathematical-physical model of the whole system using the Green function method and give an analytical expression for the thermal conductance between NPs. The effects of different parameters on the thermal transport properties of the cylinder are explored. In addition, we reveal the regulation mechanism of the rotation angle between the connecting lines of the two NPs and the cylinder's axis on RHT. This work offers potential applications for efficient energy transfer and utilization at the micro- and nanoscale.

## 2. Theoretical aspects

For illustrative purposes of this work, the geometric model is constructed as illustrated in Fig. 1. We focus on the RHT of two spherical NPs (labeled with $P_1$ and $P_2$) in the presence of an infinitely long cylinder with a radius of $R_c$. All objects are considered to be embedded in a vacuum. The RHT between two NPs can be visualized as two transport channels. Channel 1 is viewed as a direct photon-mediated heat exchange in free space, which is strongly dependent on the distance between the two interacting NPs. Channel 2 is the heat transfer assisted by cylindrical surface modes, which is attributed to the coupling of surface phonons (or electrons) with electromagnetic waves at the vacuum-cylinder interface. In addition, the two NPs of radius $R = 5$ nm are positioned at a vertical distance of $z_0$ from the top of the cylinder. The separation distance between them is denoted as $d$. The entire configuration is based on a cylindrical coordinate system $(r, \varphi, z)$. The axis of the cylinder is located on the $z$-axis. The regulation of RHT can be achieved by changing the relative position between the cylinder and the two NPs through mechanical rotation. $O'$ is the rotation point, and $L$ is the horizontal distance from $O'$ to the origin of the coordinate $O$. Here, to numerically describe the position of $O'$, we define a dimensionless factor $l$, where $l = L/d$. $\alpha$ is the rotation angle between the axis and the line connecting the two NPs. The coordinates of the two NPs ($\mathbf{r}_1$ and $\mathbf{r}_2$) can be obtained from these known parameters [see Appendix A]. When $\alpha = 0°$, the two NPs are located directly above the cylinder, and their coordinates are $\mathbf{r}_1 = (R_c+z_0, 0, 0)$ and $\mathbf{r}_2 = (R_c+z_0, 0, d)$, respectively.

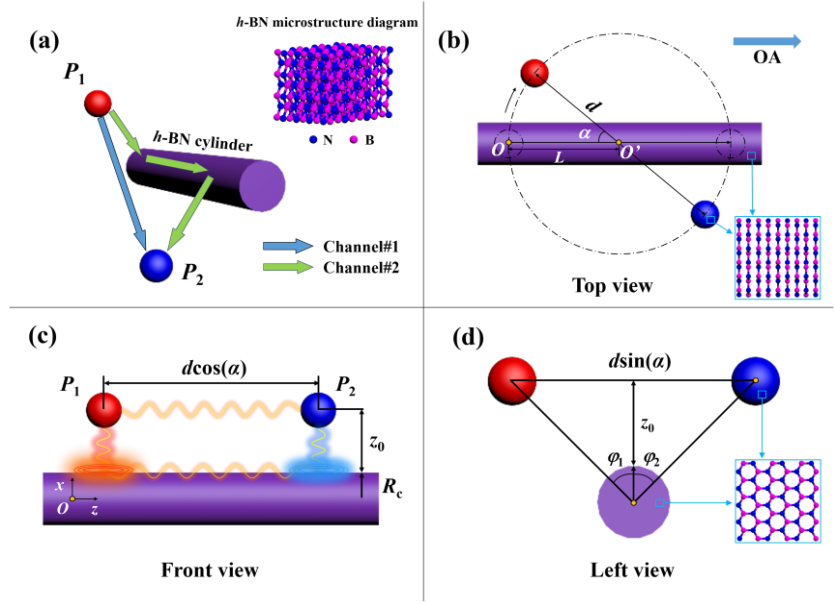

**Fig. 1**. The schematic of RHT between two NPs in the presence of an infinitely long cylinder. (a) The three-dimensional model and the microscopic structures, (b) top view, (c) front view, and (d) left view.

In this work, RHT between two NPs in the rotating system is described within the framework of fluctuational electrodynamics [28,53]. To simplify the mathematical-physics model, two NPs can be conceptualized as point-like sources. The calculation process is based on the dipole approximation method, which is valid when the length scale involved in the system is greater than three times the radius of NPs [5,29,54-57]. The temperatures of NPs $P_i$ ($i$ = 1, 2), cylinder, and the background are kept at $T_i$, $T_c$, and $T_b$, respectively. To achieve heat exchange between NPs, it is assumed that $T_1 = T + \triangle T$ ($\triangle T \ll T$) and $T_2 = T_c = T_b = T$. We give a fixed temperature $T$ = 300 K. When temperature difference $\triangle T$ tends to zero, the RHT between NPs can be described by the thermal conductance $h$, which is expressed by the Green function (GF) as [16,41]

$$h = 2\int_0^{+\infty} \frac{d\omega}{\pi} \frac{\partial \Theta(\omega,T)}{\partial T} k_0^4 \mathrm{Tr}(\overline{\overline{\chi}} \mathbb{G} \overline{\overline{\chi}} \mathbb{G}^*), \qquad (1)$$

where $\Theta(\omega,T) = \hbar\omega/[\exp(\hbar\omega/k_B T)-1]$ is the mean energy of a harmonic oscillator in thermal equilibrium at temperature $T$. $\hbar$ represents the reduced Planck constant, and $k_B$ is the Boltzmann's constant. $k_0 = \omega/c$, and $c$ is the speed of light. When neglecting the radiative correction, the dressed polarizability tensor of NPs is defined as [23,58]

$$\overline{\overline{\chi}} = \frac{1}{2i}(\overline{\overline{\alpha}} - \overline{\overline{\alpha}}^*), \qquad (2)$$

where the quasistatic polarizability tensor is given by $\overline{\overline{\alpha}} = 4\pi R^3(\overline{\overline{\varepsilon}}-\mathbf{I})(\overline{\overline{\varepsilon}}+2\mathbf{I})^{-1}$, and $\overline{\overline{\varepsilon}}$ is the permittivity tensor. $\mathbf{I}$ denotes the unit dyadic tensor.

The $\mathbb{G}$ is the dyadic Green tensor of the entire system. In the presence of the vacuum-material interface, the GF of the entire configuration can be written as

$$\mathbb{G} = \mathbb{G}_0 + \mathbb{G}_{sc}, \tag{3}$$

i.e., separated into a vacuum contribution and a scattering part. The vacuum part can be described by the free-space GF as [16,31]

$$\mathbb{G}_0 = \frac{\exp(ik_0 r)}{4\pi r} \times [(1 + \frac{ik_0 r - 1}{k_0^2 r^2})\mathbf{I} + \frac{3 - 3ik_0 r - k_0^2 r^2}{k_0^2 r^2} \hat{\mathbf{r}} \otimes \hat{\mathbf{r}}], \tag{4}$$

with the unit vector $\hat{\mathbf{r}} = \mathbf{r}/r$, $\mathbf{r}$ being the vector linking the center of NPs $P_1$ and $P_2$, and $r = |\mathbf{r}|$. The scattering contribution $\mathbb{G}_{sc}$ from the vacuum-cylinder interface can be described as [49,59]

$$\mathbb{G}_{sc} = \frac{i}{8\pi} \sum_{P,P'} \sum_{n=-\infty}^{\infty} (-1)^n \int_{-\infty}^{\infty} dk_z \mathbf{P}_{n,k_z}^{\text{out}}(\mathbf{r}) \otimes \mathbf{P}_{-n,-k_z}^{'\text{out}}(\mathbf{r}') T_{n,k_z}^{PP'}, \tag{5}$$

where $k_z$ is the z component of the wave vector in a vacuum. $n$ represents the multipole order, and $P$, $P'$ = {M, N} denote polarization (M for magnetic and N for electric). $\mathbf{M}_{n,k_z}^{\text{out}}(\mathbf{r})$ and $\mathbf{N}_{n,k_z}^{\text{out}}(\mathbf{r})$ are the outgoing waves of cylindrical harmonics as a function of $n$ and $k_z$, which can be seen in Appendix B. Based on the electromagnetic boundary conditions of three components at the vacuum-cylinder interface, the scattering matrix element $T_{n,k_z}^{PP'}$, corresponding to the scattering operator $\mathbb{T}$, depends on the radius $R_c$ and the material properties, which can be deduced as [49,51,59-61]

$$T_{n,k_z}^{MM} = -\frac{J_n(qR_c)}{H_n(qR_c)} \frac{I_1 I_4 - K^2}{I_1 I_2 - K^2}, \tag{6}$$

$$T_{n,k_z}^{NN} = -\frac{J_n(qR_c)}{H_n(qR_c)} \frac{I_2 I_3 - K^2}{I_1 I_2 - K^2}, \tag{7}$$

$$T_{n,k_z}^{NM} = T_{n,k_z}^{MN} = \frac{2i}{\pi\sqrt{\varepsilon_z}(qR_c)^2} \frac{K}{(I_1 I_2 - K^2)[H_n(qR_c)]^2}, \tag{8}$$

where $J_n$ and $H_n$ represent the Bessel function and the Hankel function of the first kind of $n$ order. $q = \sqrt{k_0^2 - k_z^2}$ is the wave vector perpendicular to the z-axis in vacuum. These intermediate variables can be represented as

$$I_1 = \frac{J_n'(q_N R_c)}{q_N R_c J_n(q_N R_c)} - \frac{1}{\varepsilon_z} \frac{H_n'(qR_c)}{qR_c H_n(qR_c)}, \tag{9}$$

$$I_2 = \frac{J_n'(q_M R_c)}{q_M R_c J_n(q_M R_c)} - \frac{H_n'(qR_c)}{qR_c H_n(qR_c)}, \tag{10}$$

$$I_3 = \frac{J'_n(q_N R_c)}{q_N R_c J_n(q_N R_c)} - \frac{1}{\varepsilon_z} \frac{J'_n(q R_c)}{q R_c J_n(q R_c)}, \tag{11}$$

$$I_4 = \frac{J'_n(q_M R_c)}{q_M R_c J_n(q_M R_c)} - \frac{J'_n(q R_c)}{q R_c J_n(q R_c)}, \tag{12}$$

$$K = \frac{n k_z c}{\sqrt{\varepsilon_z} R_c^2 \omega} \left( \frac{1}{q_M^2} - \frac{1}{q^2} \right), \tag{13}$$

where $J'_n$ and $H'_n$ are the first order derivatives of $J_n$ and $H_n$, respectively. $q_M$ and $q_N$ are the wave vectors perpendicular to the $z$-axis inside the cylinder, respectively, i.e., $q_M = \sqrt{\varepsilon_r k_0^2 - k_z^2}$ and $q_N = \sqrt{\varepsilon_z / \varepsilon_r} \sqrt{\varepsilon_r k_0^2 - k_z^2}$. We now have all the elements needed to calculate the RHT between NPs in the rotating system composed of two NPs and a uniaxial anisotropic cylinder.

## 3. Results and discussion

Based on the geometric and mathematical-physical models constructed in the previous section, the main objective of this section is to investigate the effect of the cylinder on inter-particle RHT. All material bodies are surrounded by fluctuating electromagnetic fields due to thermal and quantum fluctuations in internal current density [27,62]. The RHT depends strongly on the material properties of interacting objects. In this work, we assume that both the NPs and the cylinder are made of a layered van der Waals crystal structure (uniaxial anisotropy) called hexagonal boron nitride ($h$-BN), a two-dimensional material that supports low-loss hyperbolic phonon polaritons (HPPs) in the mid-infrared region [7,63]. In addition, recent literature has shown that cylindrical structures have low-loss propagation properties in electromagnetic energy transport [50-52]. Therefore, we consider that the combination of the two (the $h$-BN cylinder) may offer potential possibilities for enhancing and regulating long-range near-field RHT. The $h$-BN permittivity tensor can be approximated with a Drude-Lorentz model as [58,63,64]

$$\varepsilon_m = \varepsilon_{\infty,m} \left( 1 + \frac{\omega_{LO,m}^2 - \omega_{TO,m}^2}{\omega_{TO,m}^2 - i\gamma_m \omega - \omega^2} \right), \tag{14}$$

where $m = \parallel, \perp$. "$\parallel$" and "$\perp$" represent components parallel and perpendicular to the optic axis (OA), respectively. Other parameters are denoted as $\omega_{TO,\perp} = 1370$ cm$^{-1}$, $\omega_{LO,\perp} = 1610$ cm$^{-1}$, $\omega_{TO,\parallel} = 780$ cm$^{-1}$, $\omega_{LO,\parallel} = 830$ cm$^{-1}$, $\varepsilon_{\infty,\parallel} = 2.95$, $\varepsilon_{\infty,\perp} = 4.87$, $\gamma_\parallel = 4$ cm$^{-1}$, and $\gamma_\perp = 5$ cm$^{-1}$. Subsequently, we calculate the real part of the dielectric function $\overline{\overline{\varepsilon}}$ and the dressed polarizability $\overline{\overline{\chi}}$ of $h$-BN materials, as shown in Fig. 2. The types I and II are the so-called Reststrahlen bands. They lie in the ranges between the longitudinal (LO) and transverse optical (TO) phonon frequencies, where the real part of the dielectric permittivity is negative. In both ranges, *h*-BN materials can exhibit hyperbolic properties that significantly enhance the RHT between interacting objects. It can be seen that the dressed polarizability of the NPs exhibits two resonance peaks at $\omega_{np1} = 1.526 \times 10^{14}$ rad s$^{-1}$ and $\omega_{np2} = 2.908 \times 10^{14}$ rad s$^{-1}$, which arise under the condition $\varepsilon_m + 2 = 0$. The optical response of *h*-BN materials is related to the orientation of the OA [65]. In this work, it is assumed that the OA of the entire system (including two NPs and a cylinder) is along the *z*-axis. The OA of the two NPs does not change during this rotation. The dielectric tensor of the *h*-BN cylinder can expressed as $\bar{\bar{\varepsilon}} = \text{diag}(\varepsilon_r, \varepsilon_r, \varepsilon_z)$, where $\varepsilon_r = \varepsilon_\perp$ and $\varepsilon_z = \varepsilon_\parallel$.

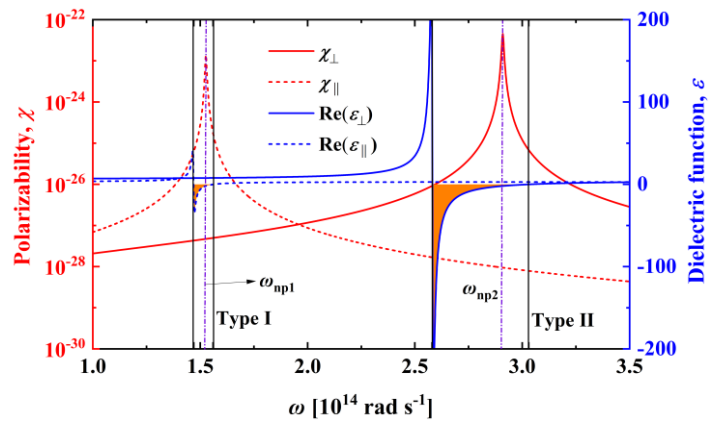

**Fig. 2**. The perpendicular and parallel components of $\bar{\bar{\chi}}$ and the real part of $\bar{\bar{\varepsilon}}$ for *h*-BN materials. The orange color highlights the hyperbolic regions.

### 3.1. Radiative thermal transport behaviors for the *h*-BN cylinder

To gain a preliminary insight of the potential of the rotating system to regulate RHT between NPs, the initial strategy we consider is to explore the thermal transport properties of the *h*-BN cylinder. According to the basic case that there are two NPs directly above the cylinder ($\alpha = 0°$), the geometric parameters ($z_0$, $d$, and $R_c$) are optimized and analyzed within the entire system, as shown in Fig. 3. To highlight the amplification of different substrates, the enhancement ratio is defined as $\Phi = h/h_0$. $h_0$ is the thermal conductance between two NPs in a vacuum. For the fixed parameters $R_c = 0.1$ μm and $z_0 = 50$ nm, the thermal conductance $h$ for different cases is calculated [see Fig. 3(a)]. The cylinders made of gold (Au) and silicon carbide (SiC) as well as a *h*-BN thin film are presented for comparison. Their material properties can be found in Appendix C. It can be seen that when the distance $d$ is small, the decay tendency of $h$ for all cases is the same as the vacuum case, i.e., $d^{-6}$. This is because for a small $d$, the coupling interaction of localized hyperbolic phonon polaritons (LHPPs) between the two *h*-BN NPs dominates RHT, and the scattering contribution from substrates is negligible. As $d$ increases, the interfacial effect of these substrates begins to play a dominant role, with different substrates producing different decreasing trends. For these two cylinders made of $h$-BN and SiC, the attenuation of $h$ gradually tends to $d^{-2}$, which is the same as the vacuum case when $d > \lambda_T$. Compared with the SiC cylinder, the $h$-BN cylinder demonstrates superior thermal transport properties. This suggests the coupling of the LHPPs excited by NPs with the hyperbolic surface mode of the $h$-BN cylinder is stronger. The reason may be that compared with the SiC cylinder, the resonant frequencies (NPs and the cylindrical surface) match better for the $h$-BN cylinder. In addition, the gold cylinder can exhibit excellent energy transfer behavior similar to that of a perfectly conducting cylinder ($|\varepsilon| \to \infty$) when $R_c = 10^{-7}$ m, which results in a higher $h$ in the presence of gold cylinders than in the presence of $h$-BN cylinders when $d = 6\text{-}100$ μm. However, this transfer behavior of gold cylinders will decay exponentially at a larger $d$ due to the presence of material loss [49-51]. Subsequently, we explore the effect of the $h$-BN thin film with a thickness of 0.2 μm ($2R_c$) on RHT. It is found that with increasing $d$, the $h$ in the presence of the thin film gradually converges to $h_0$. This phenomenon is different from that of the cylindrical structure. The $h$-BN cylinder has excellent thermal transport behaviors at large distances compared with the planar structure. The reason may be that the energy can be transferred directionally along the cylindrical surface through different guide modes (orders) [66-68]. This transfer behavior is superior to the surface modes generated by a thin film in terms of long-distance energy transfer. It is worth noting that the scattering GF of the $h$-BN thin film can be found in the literature [58]. We can see from the inset that with the increase of $d$, the enhancement ratio $\Phi$ gradually increases and eventually remains constant for the $h$-BN and SiC cylinders. Accordingly, as $d$ increases, the $\Phi$ in the presence of a gold cylinder increases gradually, while the $\Phi$ in the presence of a thin film first increases and then decreases.

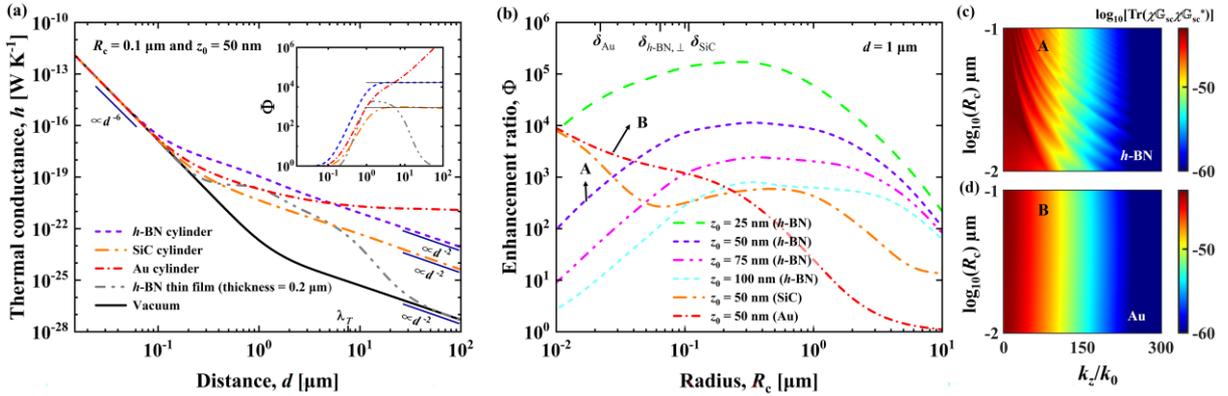

Fig. 3. Optimization of geometric parameters. (a) The thermal conductance $h$ as a function of the distance $d$ when $R_c = 0.1$ μm and $z_0 = 50$ nm. $\lambda_T$ is the thermal wavelength at temperature $T = 300$ K, which is about 7.6 μm. The inset shows the enhancement ratio $\Phi$ as a function of $d$. (b) The relationship between $\Phi$ and the cylindrical radius $R_c$ when $d = 1$ μm. $\delta_{Au}$, $\delta_{h\text{-BN},\perp}$, and $\delta_{SiC}$ are the smallest penetration depth of the cylinders made of Au, $h$-BN, and SiC, respectively, where $\delta(\omega) = c/(\operatorname{Im}\sqrt{\varepsilon}\omega)$. The trace of the GF $\operatorname{Tr}(\overline{\overline{\chi}}\mathbb{G}_{sc}\overline{\overline{\chi}}\mathbb{G}_{sc}^*)$ as a function of $R_c$ and $k_z/k_0$ when $d = 1$ μm and $\omega_{np2} = 2.908 \times 10^{14}$ rad s$^{-1}$, (c) $h$-BN cylinder and (d) Au cylinder.

Subsequently, to further explore the effect of the cylindrical size on the thermal transport behavior, the relationship between the enhancement ratio $\Phi$ and the radius $R_c$ is investigated, as shown in Fig. 3(b). At a fixed distance $d = 1$ μm, the $\Phi$ for all $h$-BN cases increases first and then decreases (non-monotonic behavior) with the increase of $R_c$. For each $z_0$, there exists an optimal radius $R_c$, which may correspond to the most excellent energy transfer performance. This phenomenon is related to the penetration depth $\delta_{h\text{-BN},\perp}$ in the radial direction of the $h$-BN cylinder. When $R_c < \delta_{h\text{-BN},\perp}$, the entire volume of the cylinder participates in the RHT between NPs, and the contribution of the $h$-BN cylinder increases with increasing radius $R_c$. While for $R_c > \delta_{h\text{-BN},\perp}$, only the thermal fluctuations on the cylindrical surface can contribute to the heat exchange. But for larger radii ($R_c \gg \delta_{h\text{-BN},\perp}$), the $h$-BN cylinder gradually tends to be a bulk structure. The coupling interaction of surface evanescent modes weakens. Therefore, the enhancement ratio $\Phi$ also decreases. It is noticed here that the contribution to RHT of the higher orders $n$ in the cylindrical Green function gradually increases with increasing $R_c$. The calculation of the $h$-BN cylinder with a large radius may be more difficult. In addition, we calculate the $\Phi$ in the presence of Au and SiC cylinders when $z_0 = 50$ nm. Both have excellent thermal transport behavior at small radii. As $R_c$ increases, their enhancement ratio is lower than that of the $h$-BN cylinder under the same $z_0$. It can be seen intuitively from Fig. 3(c) and (d) that when $R_c = 10^{-2}$-$10^{-1}$ μm, the $\text{Tr}(\bar{\bar{\chi}}\mathbb{G}_{sc}\bar{\bar{\chi}}\mathbb{G}_{sc}^*)$ in the presence of the gold cylinder gradually decreases with the increase of $R_c$. The $h$-BN cylinder shows obvious hyperbolic behaviors in the $R_c$ and $k_z/k_0$ space, and the $\text{Tr}(\bar{\bar{\chi}}\mathbb{G}_{sc}\bar{\bar{\chi}}\mathbb{G}_{sc}^*)$ in the presence of the $h$-BN cylinder increases with the increase of $R_c$. This phenomenon is consistent with the conclusion obtained in Fig. 3(b).

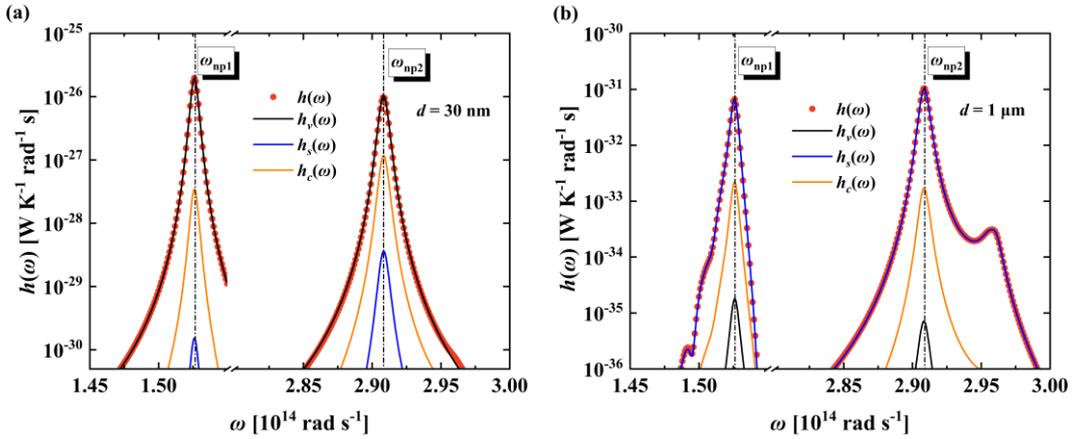

**Fig. 4**. Spectral analysis between the two NPs at distances (a) $d = 30$ nm and (b) $d = 1$ μm.

The GF within the entire system dominates the RHT between the two NPs. To gain insight into the transport properties of the $h$-BN cylinder, we expand the trace of the GF into four terms as shown in Eq. (15) [41].

$$\text{Tr}(\bar{\bar{\chi}}\mathbb{G}\bar{\bar{\chi}}\mathbb{G}^*) = \text{Tr}(\bar{\bar{\chi}}\mathbb{G}_0\bar{\bar{\chi}}\mathbb{G}_0^*) + \text{Tr}(\bar{\bar{\chi}}\mathbb{G}_{sc}\bar{\bar{\chi}}\mathbb{G}_{sc}^*) + \text{Tr}(\bar{\bar{\chi}}\mathbb{G}_0\bar{\bar{\chi}}\mathbb{G}_{sc}^*) + \text{Tr}(\bar{\bar{\chi}}\mathbb{G}_{sc}\bar{\bar{\chi}}\mathbb{G}_0^*), \tag{15}$$

where it can be divided into three parts: the vacuum part (the first term), the scattering part (the second term), and the cross part (the last two terms). It is worth noting that in terms of a uniaxial anisotropic system, the contributions from the two cross terms are numerically equal. According to the origin of the GF, the thermal conductance can be expanded as $h = h_v + h_s + h_c$, where $h_{v,s,c}$ represents the vacuum, scattering, and cross contribution, respectively. With fixed parameters ($R_c$ = 0.1 μm and $z_0$ = 50 nm), we perform spectral analysis for the cases of $d$ = 30 nm and 1 μm, respectively, as shown in Fig. 4. The RHT between NPs is mainly dominated by two resonance peaks ($\omega_{np1}$ and $\omega_{np2}$), which are determined by the material properties of $h$-BN. It can be seen from Fig. 4(a) that the vacuum contribution $h_v(\omega)$ agrees well with the total spectral thermal conductance $h(\omega)$ when $d$ = 30 nm. This proves that when the distance $d$ is small, the photon-mediated direct RHT (channel#1) plays a critical role in the heat exchange. The scattering contribution $h_s(\omega)$ from the vacuum-cylinder interface (channel#2) is negligible. When $d$ = 1 μm [see Fig. 4(b)], the propagating hyperbolic surface mode excited by the scattering GF dominates the thermal transport behavior. This provides inspiration for the twist-induced regulation of the near-field energy transport at the micro- and nanoscale system.

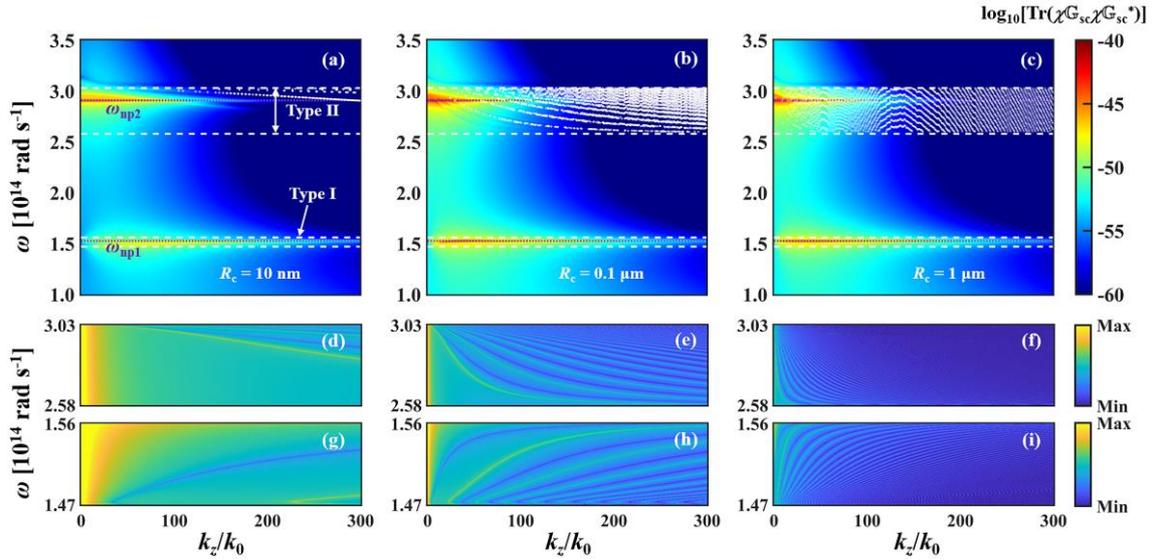

**Fig. 5.** The resonance modes provided by the $h$-BN cylinder. (a)-(c) The trace of the scattering GF $\text{Tr}(\overline{\overline{\chi}}\mathbb{G}_{sc}\overline{\overline{\chi}}\mathbb{G}_{sc}^*)$ as a function of $\omega$ and $k_z/k_0$ at $d$ = 1 μm and $z_0$ = 50 nm. The purple dotted lines represent $\omega_{np1}$ and $\omega_{np2}$, and these white dotted lines indicate the boundaries of the two reststrahlen bands. These white dots in type II are the conditions under which the dispersion equation holds when $n$ = 0. (d)-(i) The denominator of the $T^{NN}_{0,k_z}$ element as a function of $\omega$ and $k_z/k_0$ in types I and II. (d) and (g) for $R_c$ = 10 nm, (e) and (h) for 0.1 μm, as well as (f) and (i) for 1 μm. The minimum value (blue) can be considered as the dispersion curve.

The scattering contribution from the cylinder plays a pivotal role in the long-range near-field RHT between NPs. To gain insight into the underlying physical mechanism, the trace of the scattering GF $\text{Tr}(\overline{\overline{\chi}}\mathbb{G}_{sc}\overline{\overline{\chi}}\mathbb{G}_{sc}^*)$ as a function of $\omega$ and $k_z/k_0$ is investigated for different cylindrical radii $R_c$, as shown in Fig. 5. We consider that both the frequency and the wave vector are in real space. The wave vector along the z-axis $k_z$ is normalized by $k_0$. From Figs. 5(a)-(c), it can be seen that the larger values in the scattering terms are mainly distributed in the two reststrahlen bands (types I and II). These two resonance frequencies ($\omega_{np1}$ and $\omega_{np2}$) largely dominate the thermal transport behavior between NPs. The presence of a cylinder may enhance RHT by introducing an additional channel. As $R_c$ increases, the cylindrical effect gradually shifts from the volume mode to the surface mode, and the contribution from higher order $n$ in the scattering Green's function also increases. It can be seen that the trace of GF $\text{Tr}(\bar{\bar{\chi}}\bar{\bar{G}}_{sc}\bar{\bar{\chi}}\bar{\bar{G}}_{sc}^*)$ at $R_c = 0.1$ and 1 μm is numerically higher than that at $R_c = 10$ nm. In addition, the $T_{n,k_z}^{NN}$ element of the scattering operator plays a pivotal role in the energy transport process. Based on the condition that the denominator of the $T_{n,k_z}^{NN}$ is equal to zero, i.e., $I_1 I_2 - K^2 = 0$ [see Eqs. (9), (10), and (13)], we give the dispersion mode for the uniaxial anisotropic cylinder. When the order $n = 0$ (fundamental mode), the dispersion equation can be expressed as

$$\left( \frac{J_1(q_N R_c)}{q_N R_c J_0(q_N R_c)} - \frac{1}{\varepsilon_z} \frac{H_1(qR_c)}{qR_c H_0(qR_c)} \right) \left( \frac{J_1(q_M R_c)}{q_M R_c J_0(q_M R_c)} - \frac{H_1(qR_c)}{qR_c H_0(qR_c)} \right) = 0, \tag{16}$$

where the variables can be found in Sec. 2. As shown in Figs. 5(d)-(i), it is worth noting that the minimum value (that is, blue) in these figures shows the region under which the dispersion relation holds. With the increase in radius, the region where the dispersion relation is satisfied gradually increases, which may contribute to the heat exchange between NPs. Therefore, it can also be shown that the thermal transport properties of h-BN cylinders with a radius of $R_c = 0.1$ and 1 μm are better than those of a cylinder with a radius of $R_c = 10$ nm.

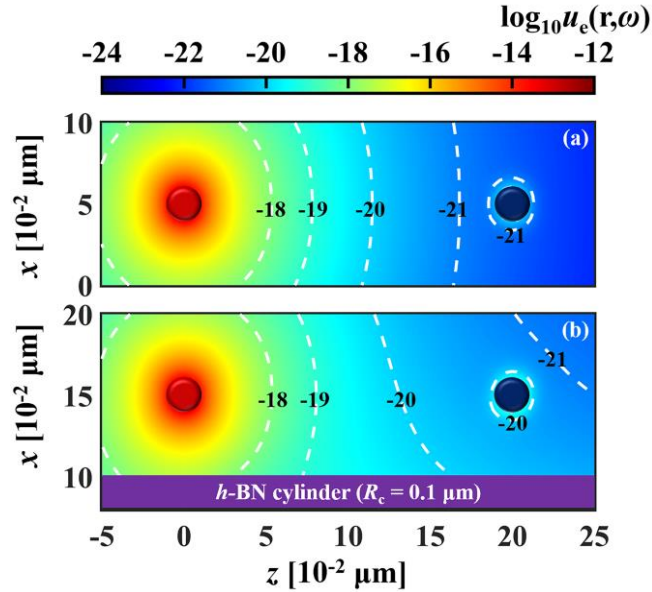

**Fig. 6.** The electric field energy density $u_e(\mathbf{r},\omega)$ distribution for the cases of $\omega_{np2}$ at $d = 0.2$ μm in the $y = 0$ plane. (a) Vacuum case and (b) the h-BN cylinder with $R_c = 0.1$ μm.

According to the material properties of h-BN, the single-frequency $\omega_{np2}$ plays a dominant role in RHT (see Fig. 2). To intuitively exhibit the thermal transport properties of the *h*-BN cylinder, the electric field energy density $u_e(\mathbf{r},\omega)$ distribution at $\omega_{np2}$ is calculated when $d = 0.2$ μm, as shown in Fig. 6. Here, the numerical model of $u_e(\mathbf{r},\omega)$ can be expressed as [31,40]

$$u_e(\mathbf{r},\omega) = (2\varepsilon_0^2/\pi\omega)\sum_j \Theta(\omega,T_j)\text{Tr}[\mathbb{Q}_{rj}\chi_j\mathbb{Q}_{rj}^*], \qquad (17)$$

where $\mathbb{Q}_{rj} = \omega^2\mu_0 \mathbf{G}^{rj}[\mathbf{I}-\mathbf{A}]^{-1}$, and $\Theta(\omega,T_j) = \hbar\omega/\exp[(\hbar\omega/k_B T_j)-1]$ represents the mean energy of the Planck oscillator at the temperature $T_j$. The energy density $u_e(\mathbf{r},\omega)$ generated by the left NP $P_1$ is much higher than that of its surroundings and decays rapidly with increasing distance from $P_1$. Compared to the vacuum case, the presence of the *h*-BN cylinder increases the region characterized by high energy density, which leads to an increase in the energy density of the right NP $P_2$. This indicates that the *h*-BN cylinder has excellent thermal transport properties.

### 3.2. Active regulation of the RHT between NPs

In the previous section, it has been demonstrated that the presence of the *h*-BN cylinder can significantly enhance RHT. Another critical issue is how to achieve active regulation of heat exchanges. Subsequently, a strategy we consider is to regulate the inter-particle RHT by changing the relative positions of the NPs and the cylinder. Initially, the thermal conductance *h* as a function of the rotation angle *α* is investigated for fixed geometrical parameters ($R_c = 0.1$ μm, $d = 1$ μm, and $z_0 = 50$ nm), as depicted in Fig. 7. $z_0$, *l*, and *d* can be found in Fig. 1. When $l = 0$, it means that the rotation point *O'* is at the position of the left NP $P_1$, while when $l = 0.5$, *O'* is located in the middle of the line connecting the two NPs. Due to the symmetry of the position, we consider only the range of the factor $l = 0\text{-}0.5$.

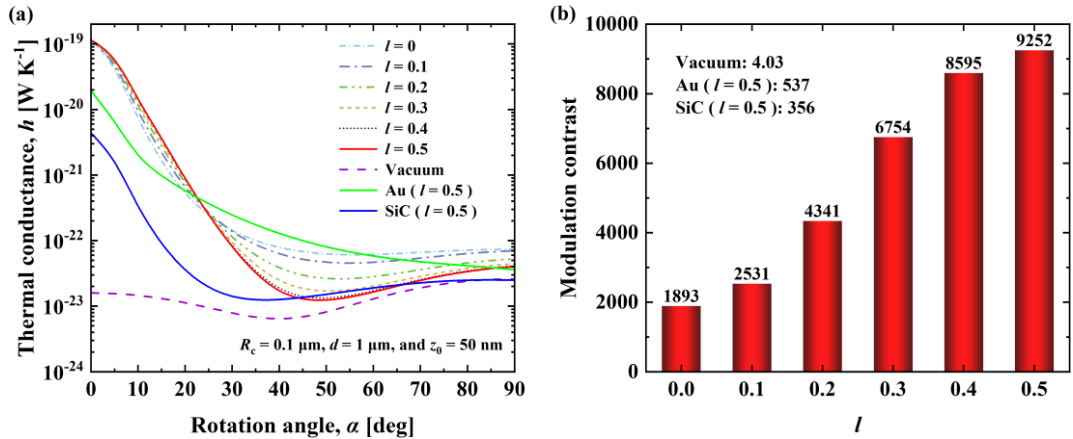

**Fig. 7.** (a) Thermal conductance *h* as a function of the rotation angle *α* for different *l*. (b) The relationship between modulation contrast and *l*.

It can be found from Fig. 7(a) that for two isolated NPs in a vacuum, the thermal conductance *h*

first decreases and then increases with the increase of α, and the maximum value occurs at α = 90°. This phenomenon may be related to the relative orientation of the two h-BN NPs. Specifically, the RHT between NPs with the same orientations tends to be higher than that with different orientations. In addition, as α increases, the h for all systems in the presence of the h-BN cylinder decreases dramatically and then increases. This is due to the twist-induced effect of the cylinder. The point of rotation O' has a significant influence on RHT. We find that when l = 0.5, the h in the presence of a SiC cylinder first decreases and then increases as α increases, and eventually converges to the vacuum state. For a gold cylinder, the h decreases gradually and above the vacuum case throughout the rotating space. This indicates that compared with h-BN and SiC cylinders, the gold cylinder plays a critical role in the whole rotating space. Figure 7(b) shows the modulation contrast (MC) at different l, which is defined as the ratio of maximum thermal conductance to minimum value. It can be seen that the MC gradually increases as l increases. The maximum value is 9252, which occurs at l = 0.5. In contrast, the MC in a vacuum is only 4.03. The MC in the presence of SiC and gold cylinders is 537 and 356, respectively. This demonstrates that active control and regulation of the RHT can be achieved by changing the position of the NPs relative to the cylinder. Compared with other cases, the h-BN cylinder offers excellent tunability in terms of energy transfer.

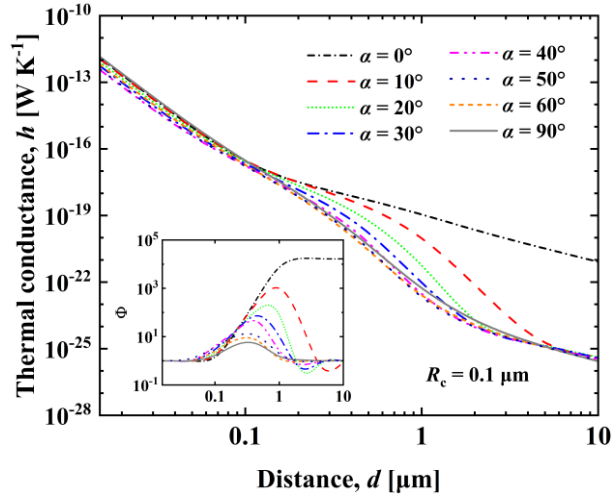

**Fig. 8.** For different rotation angles α, thermal conductance h as a function of d when l = 0.5. The inset shows the relationship between the enhancement ratio Φ and the distance d.4

Subsequently, we fix the rotation point O' at the middle of the line connecting the two NPs, i.e., l = 0.5. The thermal conductance h as a function of the inter-particle distance d is investigated, as illustrated in Fig. 8. When d is small (less than 0.1 μm), the difference in the thermal conductance for different rotation angles α is mainly due to the different relative orientations of the two NPs. When d > 0.1 μm (approximately), the contribution from the cylinder gradually comes to the fore. With increasing d, the thermal conductance h for different α gradually converges to their corresponding vacuum states, except for the anomalous behavior at α = 0°. The larger the rotation angle α is, the faster the convergence is. This phenomenon may be attributed to the near-field interaction between the two NPs and the cylinder. When $\alpha = 0°$, the vertical distance $z_0$ between the NPs and the cylinder does not change with the increase of $d$, resulting in a strong near-field effect. The cylinder shows excellent long-range thermal transport characteristics. However, with the augment of $\alpha$, the near-field effect weakens dramatically with the increase of $z_0$, and eventually be neglected. Therefore, when $\alpha > 0°$, the thermal conductance close to the vacuum cases with increasing $d$. It is worth noting that the thermal conductance at $\alpha = 90°$ is higher at some positions than that at angles smaller than it ($\alpha = 40$-$60°$). This is due to the combined effect of the relative orientations of the two NPs and the surface modes of the cylinder. It can be seen from the inset that with the increase of $\alpha$, the maximum enhancement ratio $\Phi$ gradually decreases, which also indicates that the thermal transport behavior of the $h$-BN cylinder is gradually weakened.

To deeply explain the above phenomena, the spectral conductance of scattering contributions $h_s$ within types I and II are explored, respectively, as shown in Figs. 9(a) and 9(b). The rotation angle does not change the positions of the two resonance peaks. The heat exchange is dominated by the resonance thermal transport behavior. The thermal conductance $h$ of the two peaks essentially decreases as the rotation angle increases. When $\alpha \geqslant 30°$, $\omega_{np2}$ is numerically higher than $\omega_{np1}$. This means that when $\alpha$ is large, the RHT is mainly dominated by the resonance peak $\omega_{np2}$, which occurs within the type II. Subsequently, the ratio of the scattering contribution to vacuum $h_s(\omega)/h_v(\omega)$ within type II is plotted, as shown in Fig. 9(c). When $\alpha$ is large, $\log_{10}[h_s(\omega)/h_v(\omega)]$ is less than zero at $\omega_{np2}$. This means that the scattering contribution to the RHT is less than the photon-mediated vacuum contribution. As $\alpha$ increases, the ability of the cylinder to regulate and control RHT decreases, and the interaction of LHPPs induced by NPs plays a pivotal role in heat exchanges. At this point, the RHT between the two NPs depends strongly on their relative orientations.

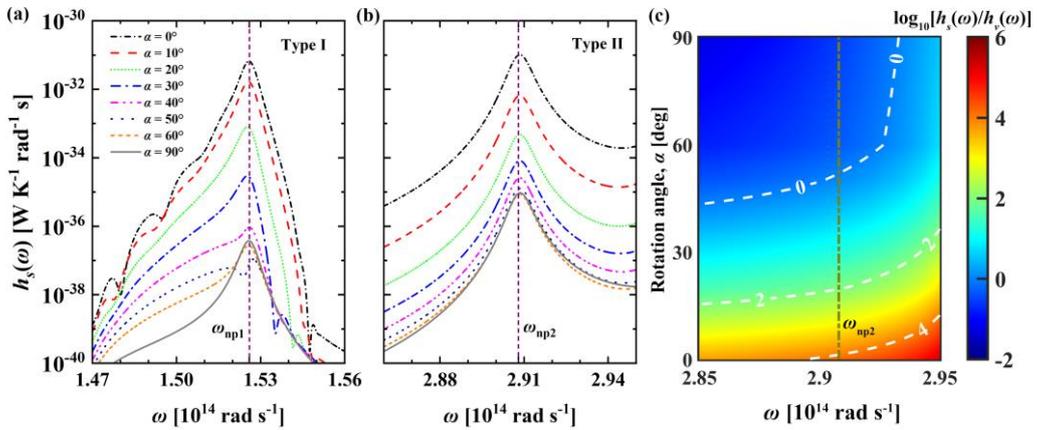

**Fig. 9.** Spectral analysis at different rotation angles $\alpha$ for (a) type I and (b) type II when $d = 1$ μm. (c) $\log_{10}[h_s(\omega)/h_v(\omega)]$ as a function of $\alpha$ and frequency $\omega$ within type II.

To visualize the effect of the relative positions of the NPs and cylinder on the thermal transport properties, we explore the electric field energy density distribution in the $x = 0.15$ μm plane for the rotation angles $α = 0, 30, 50$, and $90°$, respectively, as shown in Fig. 10. It can be seen that the energy produced by $P_1$ is much higher than that in its neighborhood and decays rapidly with increasing distance from $P_1$. The presence of the cylinder can provide an additional channel for energy transport. When $α = 0°$, the coupling effect of the hyperbolic surface modes is stronger, resulting in a higher energy for $P_2$. As $α$ increases, the distance between the two NPs and the cylinder increases, and the energy transferring along the cylindrical channel decreases, thus leading to the energy density of $P_2$ decreasing. As $α$ continues to enlarge, the energy density at $P_2$ remains almost constant, which means that the regulating effect of the cylinder disappears and the RHT between two NPs depends on the vacuum contribution. Therefore, it is also intuitively confirmed that the RHT between NPs can be regulated by changing the relative position of the NPs and the $h$-BN cylinder.

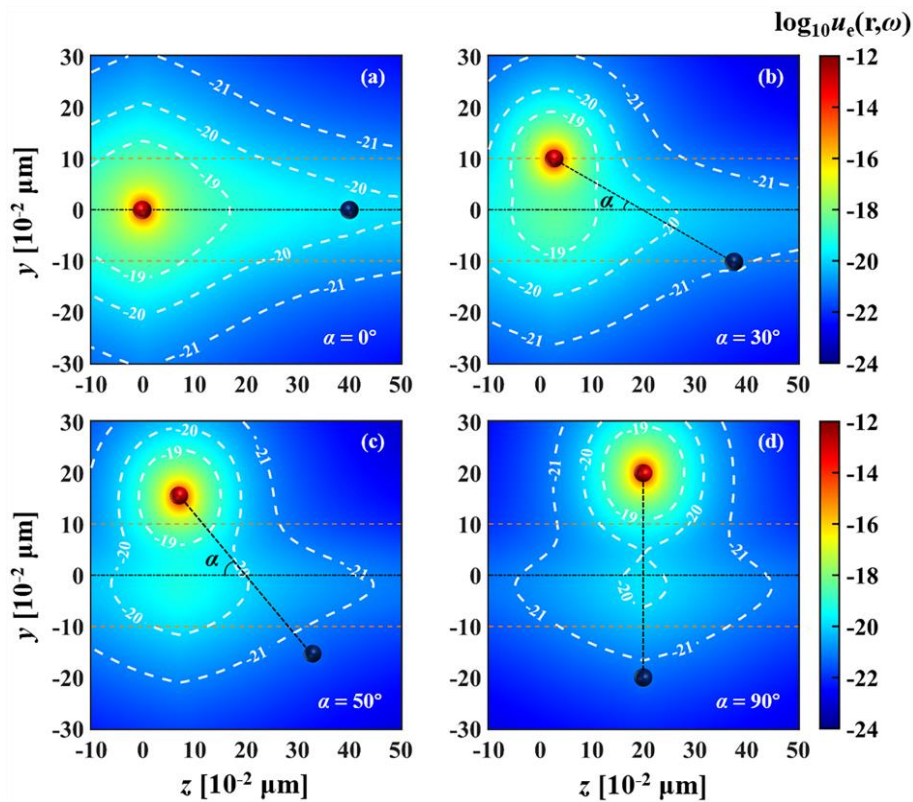

**Fig. 10**. The electric field energy density $u_e(\mathbf{r}, ω)$ distribution for the cases of $ω_{np2}$ at $d = 0.4$ μm in the plane of $x = 0.15$ μm when $α$ is equal to (a) 0, (b) 30, (c) 50, and (d) 90 degrees. The three horizontal lines represent the boundary line and the axis of the cylinder.

We have demonstrated the ability to regulate and control the RHT between NPs by changing the relative positions of the NPs to the cylinder (the rotation angle). From the perspective of universality, we subsequently explore the effect of the cylindrical geometry on the tunability of energy transfer. Here, the strategy is to investigate the thermal conductance $h$ as a function of the rotation angle $α$ for different $R_c/d$, as shown in Fig. 11. Without loss of generality, the distance is fixed at $d = 1$ μm. We can see that when $R_c/d = 0.01$, the thermal conductance $h$ decays sharply with the increase of $α$ and then approaches the vacuum case at around $\alpha = 15°$. The MC is approximately 249. When $R_c/d = 0.1$, the thermal conductance $h$ first increases and then decreases as $\alpha$ increases. The minimum value occurs at $\alpha = 50°$. Compared to the vacuum case, the $h$ in the presence of a cylinder $R_c/d = 0.1$ shows a global enhancement in the rotating space, but when $\alpha > 50°$, the enhancement is only 1-2 times, which is not significant. The MC can reach 9252, indicating that the cylinder exhibits excellent performance in regulating RHT. In addition, when $R_c/d = 0.5$ and 1, the high-order contribution from the cylinder gradually increases. With the increase of $\alpha$, the $h$ first decreases and then almost remains unchanged, higher than other cases under the same condition. The reason for this phenomenon may be that when the diameter of the cylinder ($2R_c$) is equal to or greater than the distance $d$, i.e., $R_c/d \geq 0.5$, the scattering contribution from the cylinder strongly dominates the thermal transport behavior. The change in the rotation angle $\alpha$ has less effect on RHT. Therefore, the rotating system is closely related to the diameter and the distance between the two NPs. When the diameter is slightly smaller than the distance, the $h$-BN cylinder exhibits excellent tunability in the RHT between the two interacting NPs.

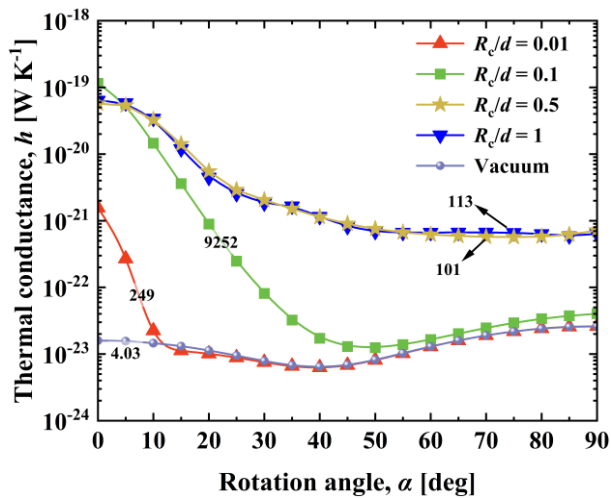

**Fig. 11**. Thermal conductance $h$ as a function of the rotation angle $\alpha$ for different $R_c/d$. The number on each curve is the modulation contrast for their corresponding case.

## 4. Conclusions

In conclusions, we present a twist-induced near-field radiative thermal regulator using a uniaxial anisotropic $h$-BN cylinder. We have explored the radiative thermal transport properties of the $h$-BN cylinder with different parameters. The results show that when two NPs are located directly above the cylinder ($\alpha = 0°$), energy can be directionally transmitted along the cylindrical channel in the form of low-loss surface waves, which significantly enhances the RHT between the NPs at large spacing. We use the spectral analysis, the trace of the Green function, and the electric field energy density distribution to analyze the phenomenon of cylinder-mediated energy transport. In addition, the RHT between NPs can be regulated by actively manipulating the excitation of cylindrical surface modes through mechanical rotation. When the rotation point is located in the middle of the line connecting the two NPs, the modulation contrast can reach 9252, which is higher than that of other cases. The rotating system is closely related to the diameter of the cylinder and the distance between NPs. When the diameter is less than the distance, the *h*-BN cylinder exhibits excellent tunability in the heat exchange. This work may provide the possibility for actively regulating and controlling near-field RHT between arbitrary objects based on cylindrical waveguides.

## CRediT authorship contribution statement

**Jian-You Wang:** Conceptualization, Methodology, Writing – original draft, Writing – review & editing, Investigation. **Yong Zhang:** Methodology, Writing – review & editing, Investigation, Supervision. **Xiao-Ping Luo:** Supervision, Investigation. **Mauro Antezza:** Writing – review & editing, Investigation. **Hong-Liang Yi:** Supervision, Investigation.

## Date availability

Data will be made available on request.

## Acknowledgements

This work was supported by the National Natural Science Foundation of China (Grants No. 52076056 and No. U22A20210).

## Appendix A: The coordinates of the two nanoparticles

Based on the known parameters in Fig. 1 of the main text, we can obtain the coordinates of the two NPs [$\mathbf{r_1} = (r_1, \varphi_1, z_1)$ and $\mathbf{r_2} = (r_2, \varphi_2, z_2)$], which are described as

$$\begin{aligned} r_1 &= \sqrt{(z_0 + R_c)^2 + [d \times \sin(\alpha) \times l]^2} \\ \varphi_1 &= -\arctan\left[\frac{d \times \sin(\alpha) \times l}{z_0 + R_c}\right] \\ z_1 &= d \times l \times [1 - \cos(\alpha)] \end{aligned} \quad , \tag{A1}$$

and

$$\begin{aligned} r_2 &= \sqrt{(z_0 + R_c)^2 + [d \times \sin(\alpha) \times (1-l)]^2} \\ \varphi_2 &= \arctan\left[\frac{d \times \sin(\alpha) \times (1-l)}{z_0 + R_c}\right] \\ z_2 &= d \times [l + (1-l) \times \cos(\alpha)] \end{aligned} \quad . \tag{A2}$$

According to the symmetry of the geometric system, the rotation angle $\alpha$ varies from 0 to 90°, and the dimensionless factor $l$ ranges from 0 to 0.5.

## Appendix B: The outgoing waves of cylindrical harmonics

According to the scattering theory of electromagnetic waves, the outgoing solution of the scatter wave equation in cylindrical coordinates $(r, \varphi, z)$ can be described as [69]

$$\psi_n(q, k_z, \mathbf{r}) = H_n(qr) e^{ik_z z + in\varphi}, \tag{B1}$$

where $H_n$ is the Hankel function of the first kind. The vector cylindrical wave functions are solved by $\mathbf{M}_{n,k_z}^{out}(\mathbf{r}) = \nabla \times [\psi_n(q, k_z, \mathbf{r}) \mathbf{e}_z]$ and $\mathbf{N}_{n,k_z}^{out}(\mathbf{r}) = \nabla \times \nabla \times [\psi_n(q, k_z, \mathbf{r}) \mathbf{e}_z]/k_0$. The magnetic multipole (TE) $\mathbf{M}_{n,k_z}^{out}(\mathbf{r})$ and electric multipole (TM) $\mathbf{N}_{n,k_z}^{out}(\mathbf{r})$ waves in component form can be written respectively as [49,51,59]

$$\mathbf{M}_{n,k_z}^{out}(\mathbf{r}) = [\frac{in}{qr} H_n(qr) \mathbf{e}_r - H_n'(qr) \mathbf{e}_\varphi] e^{ik_z z + in\varphi}, \tag{B2}$$

$$\mathbf{N}_{n,k_z}^{out}(\mathbf{r}) = \frac{1}{k_0}[ik_z H_n'(qr) \mathbf{e}_r - \frac{nk_z}{qr} H_n(qr) \mathbf{e}_\varphi + q H_n(qr) \mathbf{e}_z] e^{ik_z z + in\varphi}, \tag{B3}$$

where $k_z$ and $q$ are the wave vectors parallel and perpendicular to the cylinder $z$-axis.

## Appendix C: Physical properties of silicon carbide and gold

Silicon carbide (SiC) is a typical polar dielectric material with a wide range of applications in high-power optoelectronic semiconductors. Its dielectric function can be described by the Drude-Lorentz model [2,38,49]

$$\varepsilon_{SiC}(\omega) = \varepsilon_\infty \frac{\omega_L^2 - \omega^2 - i\Gamma\omega}{\omega_T^2 - \omega^2 - i\Gamma\omega}, \tag{C1}$$

with $\varepsilon_\infty = 6.7$, $\omega_L = 1.83 \times 10^{14}$ rad s$^{-1}$, $\omega_T = 1.49 \times 10^{14}$ rad s$^{-1}$, and $\Gamma = 8.97 \times 10^{11}$ rad s$^{-1}$. In addition, the dielectric response of gold (Au) materials can be expressed by Drude model as [38,51]

$$\varepsilon_{Au}(\omega) = 1 - \frac{\omega_p^2}{\omega(\omega + i\omega_\tau)}, \tag{C2}$$

with $\omega_p = 1.37 \times 10^{16}$ rad s$^{-1}$ and $\omega_\tau = 4.06 \times 10^{13}$ rad s$^{-1}$.